\documentstyle[psfig,fixup]{mn}

\def\ab{a_{\rm b}}
\def\ah{a_{\rm h}}
\def\th{t_{\rm h}}
\def\agr{a_{\rm gr}}
\def\vgr{v_{\rm gr}}
\def\tgr{t_{\rm gr}}
\def\vej{v_{\rm ej}}
\def\vf{v_{\rm f}}
\def\rc{r_{\rm c}}
\def\rw{r_{\rm w}}
\def\Mc{M_{\rm c}}

\def\bmax{b_{\rm max}}
\def\vbin{V_{\rm bin}}

\def\pc{\,{\rm pc}}

\def\yr{\,{\rm yr}}

\def\Cav{\langle C\rangle}
\def\Fej{F_{\rm ej}}
\def\Jgr{J_{\rm gr}}
\def\KCh{K_{\rm Ch}}
\def\M12{M_{12}}
\def\Mej{M_{\rm ej}}

\def\Sbin{\Sigma_{\rm bin}}
\def\Scap{\Sigma_{\rm cap}}

\begin{document}

\title[Massive Black Hole Binaries]{The dynamical evolution of massive black
hole binaries - I. Hardening in a fixed stellar background.}
\author[Gerald D.\ Quinlan]{Gerald D.\ Quinlan\\
Lick Observatory, University of California, Santa Cruz CA 95060 \\
Dept.\ of Physics and Astronomy, Rutgers University, PO Box 849, Piscataway
NJ 08855$^*$\\
$^*$Present address}
\date{17 January 1996}

\maketitle
\begin{abstract}
The stellar ejection rate and the rates of change of the binary semimajor
axis and eccentricity are derived from scattering experiments for the
restricted three-body problem.  They are used to study the evolution of
binaries in simple models for galactic nuclei, starting soon after the black
holes become bound and continuing until the evolution is dominated by the
emission of gravitational radiation, or until the ejected mass is too large
for the galaxy to be considered fixed.  The eccentricity growth is found to
be unimportant unless the binary forms with a large eccentricity.  The
scattering results are compared with predictions from Chandrasekhar's
dynamical-friction formula and with previous work on the capture and
scattering of comets by planetary systems.  They suggest that a binary with
masses $m_1\geq m_2$ should not be considered hard until its orbital
velocity exceeds the background velocity dispersion by a factor that scales
as $(1+m_1/m_2)^{1/2}$.
\end{abstract}

\section{Introduction}

The existence of massive black hole (BH) binaries follows from two
widely-accepted assumptions: that galaxies merge with other galaxies, and
that many galaxies contain massive BHs.  For if two BHs enter the core of a
merged galaxy, dynamical friction drags them to the center where they form a
binary.  The subsequent evolution was first outlined by Begelman, Blandford,
and Rees (1980, hereafter BBR).  Initially the binary hardens (i.e.\ its
separation shrinks) because of the interaction between the BHs and all the
stars in the galaxy core.  But that is ineffective once the BHs become close
because distant stars perturb the binary's center of mass but not its
semimajor axis.  The binary then hardens by giving kinetic energy to stars
that pass in its immediate vicinity; a hard binary can eject stars out of
the core at high velocity.  If there are enough stars for the hardening to
continue (and gas accretion onto the BHs can help), eventually the BHs merge
through the emission of gravitational radiation; otherwise the hardening
stalls and the binary survives for the lifetime of the galaxy.

Whether a binary merges or survives and how long it spends in each stage of
the evolution are questions relevant to a number of problems in
extragalactic astronomy.  Their answers would help us predict the total BH
merger rate and whether it is high enough for us to detect the resulting
gravitational waves (e.g.\ Thorne 1992, Haehnelt 1994).  They would help us
assess BH-binary models for the bending and apparent precession of radio
jets from active galactic nuclei, first proposed by BBR.  And they would
tell us what to expect if three or more BHs enter the core of a galaxy,
which can happen if the BHs are dragged in from the galaxy's halo or if the
galaxy undergoes multiple mergers with other galaxies containing BHs.  If
the first binary merges fast it can form a binary with a third BH, and once
that merges it can form a binary with a fourth, and so on, leading to a
massive central BH; but if the first binary still exists when a third BH
enters then one or all three of the BHs can be ejected in a sling-shot
interaction.  Arguments like these can set limits on massive BHs as
dark-matter candidates for galactic halos (see Hut and Rees 1992, Xu and
Ostriker 1994 for conflicting limits for our Galaxy).

Another question is what a binary merger does to the surrounding galaxy,
i.e.\ what observable signature it leaves.  Mass ejection during the
evolution should reduce a galaxy's central density and expand its core
(BBR). Ebisuzaki, Makino, and Okumura (1991) have proposed this as an
explanation for why large elliptical galaxies have lower central densities
and weaker density cusps than small ellipticals (e.g.\ Kormendy et al.\
1994).

We are far from having precise answers to any of these questions.  BBR gave
a range of merger times for one typical example that spanned three orders of
magnitude because of the uncertain influence that mass ejection has on the
hardening rate.  Fukushige, Ebisuzaki and Makino (1992) have argued that
dynamical friction causes a binary to become highly eccentric and that this
greatly reduces the merger time because gravitational radiation then becomes
important early in the evolution.  Although their arguments are not
convincing, they have called attention to the eccentricity growth and our
ignorance of its correct description.  There are uncertainties in how the
hardening rate depends on the ratio of the two BH masses, in when a binary
makes the transition from soft to hard, and even in what the words soft and
hard should mean in this context.  And our knowledge of how a binary merger
changes a galaxy is based on back-of-the-envelope estimates and simple
N-body experiments with unrealistic galaxy models.

The first step towards resolving these questions is to understand how a
massive binary evolves in fixed stellar background.  Consider a binary with
masses $m_1\geq m_2$ and semimajor axis $a$ in an isotropic background of
stars of mass $m_*\ll m_2$, density $\rho$, and one-dimensional velocity
dispersion $\sigma$; let $\M12$ and $\mu$ denote the total and reduced
binary mass:
\begin{equation} \label{eq-mdef}
   \M12 = m_1 + m_2,\qquad \mu = m_1m_2/\M12.
\end{equation}
The binary evolution and its effect on the galaxy are determined by three
dimensionless quantities: the hardening rate
\begin{equation} \label{eq-Hdef}
   H = {\sigma\over G\rho}{d\over dt}\left(1\over a\right),
\end{equation}
the mass ejection rate (where $\Mej$ is the stellar mass that the binary has
ejected from the galaxy core)
\begin{equation} \label{eq-Jdef}
   J = {1\over\M12}{d\Mej\over d\ln(1/a)},
\end{equation}
and the eccentricity growth rate
\begin{equation} \label{eq-Kdef}
   K = {de\over d\ln(1/a)}.
\end{equation}
The quantities $H$, $J$, and $K$ can be found from scattering experiments that
treat the star-binary encounters one at a time. Analytic approximations such
as the impulse approximation are helpful during the early stages of the
evolution (Gould 1991), but not once the binary becomes hard.

Most published scattering experiments assume the binaries and stars to have
equal or nearly equal masses (see Heggie 1988 for a review).  There are some
exceptions.  Roos (1981) performed scattering experiments for the restricted
three-body problem to study the evolution of hard, massive BH binaries, and
used them to correct a misplaced factor of $m_1/m_2$ in the BBR hardening
rate.  He tried to measure $K$ but his statistics were too poor to give
definite conclusions (only 500 orbits per measurement).  Hills (1983a) used
the general three-body problem to study interactions between a massive
binary and low-mass intruders ($m_*/m_2=0.01$).  He gave results for the
hardening rate for a wide range of mass ratios ($m_1/m_2=1$ -- 300), but
like Roos he considered only very hard binaries.  Mikkola and Valtonen
(1992) used the restricted three-body problem to measure $H$ and $K$ for
equal-mass BH binaries with varying degrees of hardness. Their measurements
are accurate for hard binaries but have large error bars for binaries that
are not hard.

If $m_1\gg m_2$ then the interaction between a star and a BH binary is
similar to the interaction between a comet and a planet orbiting a star.
Although scattering experiments are used to study cometary dynamics (see
Fernandez 1993 for a review) they are not of much help for our questions
about BH binaries, partly because they often consider only one mass ratio
(for the Sun-Jupiter system), but mostly because they are used to answer
different questions, such as the cross section for the capture of
interstellar comets, or for the conversion of long-period comets to
short-period comets, the survival probability of comets once they are
captured, and how all these depend on the comet's inclination.  There is
nevertheless some overlap between the two problems.

The goal of this paper is to present accurate measurements of $H$, $J$, and
$K$ over the range of parameters of interest for the BH-binary problem,
including the dependence on the mass ratio, eccentricity, and degree of
hardness of the binary.  Other quantities to be studied include the cross
section for a binary to capture stars into bound orbits, for close
encounters between stars and the binary members, and the distribution of
velocities with which stars are expelled from the binary.  These add to our
understanding of $H$, $J$, and $K$, and are needed by themselves for some
applications.  The results will be presented in a model-independent way so
that they can be applied to any problem with $m_*\ll m_2\leq m_1$.

Once $H$, $J$, and $K$ have been measured they can be used to study the
evolution of binaries in fixed galaxy models.  That will be done here for
some simple models.  If the BHs are large they will of course eject too much
mass from the galaxy core for it to be considered fixed.  But the results
will still be valid during the early stages of the evolution.  And they will
be helpful even in the later stages, because we can imagine at any instant
that the binary is embedded in a fixed background whose properties are those
of the galaxy at that instant.  The self-consistent evolution of a massive
binary in a realistic galaxy model and the changes this induces in the model
are best studied by large N-body experiments.  That will be deferred to
paper II, along with a discussion of what both papers imply for the
astronomical questions mentioned above (Quinlan and Hernquist, in
preparation).

\section{Computational method}

\subsection{Derivation of results from the restricted three-body problem}

We treat the star as a massless test particle moving in the potential of the
two BHs.  From the changes in the star's energy and angular momentum per
unit mass, $\Delta E_*$ and $\Delta L_*$, we infer the corresponding changes
$\Delta E$ and $\Delta L$ that the binary would have suffered if the star
had been given a small but nonzero mass. The three bodies are treated as
point masses and gravitational radiation is ignored.

In a real galaxy stars approach the binary with a wide distribution of
velocities at any given time. But the scattering experiments are easiest to
perform if the stars all start from the same velocity $v$ at a large
separation from the binary (the initial velocity, or the velocity at
infinity). In that case we write
\begin{equation} \label{eq-H1def}
   {d\over dt}\left(1\over a\right) = {G\rho\over v}H_1,
\end{equation}
where the ``1'' indicates that the stars all have a single velocity $v$.
The hardening rate for a Maxwellian velocity distribution is then
\begin{equation} \label{eq-Hmb}
   H(\sigma) = \int_0^\infty\!\! dv\,\,4\pi v^2
               f(v,\sigma)\,{\sigma\over v}\, H_1(v) ,
\end{equation}
where
\begin{equation} \label{eq-fmb}
   f(v,\sigma) = {1\over(2\pi\sigma^2)^{3/2}}\exp(-v^2/2\sigma^2).
\end{equation}

The hardening rate is derived from the average energy change for stars that
scatter off the binary. We define a dimensionless energy change $C$ by
(Hills 1983a)
\begin{equation} \label{eq-Cdef}
   C = {\M12 \over 2m_*}{\Delta E\over E} = {a\Delta E_*\over G\mu}.
\end{equation}
This must be averaged over all angular variables describing the binary's
orientation and phase, to give $\Cav$, and then integrated over all impact
parameters.  The averaging and integrating are done in a Monte Carlo fashion
by picking orbits from suitable distributions. We sometimes describe an
orbit by its impact parameter $b$, the distance at which it would pass the
binary if it felt no attraction, and sometimes by its pericenter distance
$r_p$, the distance if it is attracted by a point mass $\M12$. The two are
related by gravitational focusing:
\begin{equation} \label{eq-brp}
   b^2 = r_p^2\left(1 + {2G\M12\over r_pv^2}\right) .
\end{equation}
We also define a dimensionless impact parameter $x$ by
\begin{equation} \label{eq-xdef}
   x = b/b_0 ,\qquad b_0^2 = 2G\M12 a/ v^2 ;
\end{equation}
$b_0$ is approximately the
impact parameter corresponding to $r_p=a$ if gravitational focusing is
important.  With this notation we can write 
\begin{equation} \label{eq-H1}
   H_1 = 8\pi I_x(C) =  8\pi\int_0^\infty\!\! dx\,\,x\Cav ,
\end{equation}
where the second equality defines the operator $I_x$.

The derivation of the eccentricity growth rate is similar. The change to the
binary's eccentricity from a single scattering event is, if the change is
small and the binary's orbital angular momentum points in the $z$ direction,
\begin{equation} \label{eq-deda}
   {\Delta e\over\Delta\ln(1/a)} = {(1-e^2)\over 2e}\left[
   -2\left(\Delta L_z\over L_z\right) \left(E\over\Delta E\right) 
   - 1 \right] .
\end{equation}
We define a dimensionless angular-momentum change $B$ by
\begin{equation} \label{eq-Bdef}
   B = - {\M12 \over m_*}{\Delta L_z\over L_z} = 
         {\M12 \over \mu}
       { \Delta L_{*,z} \over \left[G\M12 a(1-e^2)\right]^{1/2} }, 
\end{equation}
and can then write
\begin{equation} \label{eq-K1}
   K_1 = {(1-e^2)\over 2e}\left[I_x(B-C)\over I_x(C)\right] ,
\end{equation}
where the ``1'' has the same meaning as before.  The derivation of $K$ from
$K_1$ will be described later.

\subsection{The scattering experiments}


Each scattering experiment requires five uniformly-distributed random
numbers (four if the binary is circular): one for the square of the impact
parameter (in some range [0,$\bmax^2$]), and four to fix the binary's
orientation and phase: the cosine of the inclination ([-1,1]), the longitude
of ascending node ([0,$2\pi$]), the argument of pericenter ([0,$2\pi$]), and
the mean anomaly at some fixed time ([0,$2\pi$]).  The numbers are chosen
with the quasi-random number generator {\tt sobseq} of Press et al.~(1992).

The range [0,$\bmax$] for impact parameters is split into five intervals
corresponding to ranges in scaled pericenter distance $r_p/a$ of [0,1],
[1,2], [2,4], [4,8], and [8,16].  Each output quantity is measured in a
number of steps. On the first step the program spends short but equal
amounts of cpu time picking orbits from the five intervals. On each
successive step the program doubles the cpu time and adjusts its strategy so
that the time it spends on each interval is proportional to the uncertainty
that interval contributes to the quantity being measured. Once the
uncertainty is reduced to an acceptable level, or the cpu time exceeds some
maximum allowed value, the results from all five intervals are combined with
appropriate weights for a distribution uniform in $b^2$. For $H$, $J$, and
$K$ the last three intervals contribute little because the changes in energy
and angular momentum fall off rapidly with increasing impact parameter.

The coordinates are chosen so that the binary's center of mass is at the
origin and the star starts at infinity with $(x,y,z)=(b,0,\infty)$ and
$(v_x,v_y,v_z)=(0,0,-v)$. The star is moved from $r=\infty$ to $r=50a$ along
a Keplerian orbit about a point mass $\M12$ at the origin.  The numerical
integration starts at $r=50a$.


The orbits are integrated in double precision with an explicit, embedded
Runge-Kutta method of order (7)8: the program {\tt dopri8} of Hairer,
Norsett, and Wanner (1987).  The program adjusts the integration stepsize to
keep the fractional error per step in the position and velocity below some
level $\epsilon$, which was set to $10^{-9}$.  With this choice the change
in a star's Jacobi constant for a circular binary is at most
$10^{-6}G\M12/a$ and often much smaller.  The forces from the BHs are not
softened.

Some integrations are time consuming because the star gets captured into a
weakly-bound orbit and makes many revolutions before it is expelled. The
integration stepsize is a small fraction of the binary's period even if the
period of the star about the binary is much longer.  The following
approximation expedites those experiments.  If a captured star moves further
than $r_k=a(10^{10}m_2/m_1)^{1/4}$ from the binary, the binary is replaced
by a point mass $\M12$ and the star is moved along a Keplerian ellipse until
it returns inside $r_k$, when the forces from the BHs are reintroduced with
the correct orbital phase.  The $(m_2/m_1)^{1/4}$ mass scaling makes the
quadrupole force from the binary at $r_k$ about $10^{-10}G\M12/a^2$,
independent of $m_2/m_1$.

Orbits that get captured for long times tend to be highly chaotic. The
integration for any particular orbit of that type is difficult to justify
because a small change to the integration procedure can make a big change to
the outcome. But the average results derived from a large number of
integrations can be correct even if the individual integrations are not;
that is suggested by shadowing lemmas that have been proved for simple
chaotic systems.  The average results presented here do not change
noticeably if $\epsilon$ is raised or lowered by a factor of 100, even
though some orbits undergo big changes.


An integration is stopped when the star leaves the sphere $r=50a$ with
positive energy.  The average results are not sensitive to the location of
this sphere provided that it is at least 10--15 times larger than $a$.  Once
the integration stops the program records the changes to the star's energy
and angular momentum, the minimum separations between the star and the two
BHs, and between the star and the binary's center of mass, the integration
time, the number of integration steps, and the number of times the star's
radius passed through a minimum (which, if greater than one, gives the
number of revolutions that a captured orbit made).  An orbit that does not
get captured typically takes a few hundred to a few thousand integration
steps. Captured orbits can take much longer.  If an integration lasts for
more than $10^6$ steps it is abandoned.  The fraction of abandoned
integrations is the largest for hard binaries with $m_1/m_2\gg1$, but even
for those it is less than 0.1\%.


The error in any average quantity has a systematic component and a
statistical component.  Systematic errors arise, for example, because of
errors in the numerical integration, because integrations are abandoned if
they take too long, because the program imposes a maximum impact parameter
$\bmax$, and because the orbits start and end at $r=50a$ instead of at
$r=\infty$.  But none of these is large: the total systematic error is
usually much smaller than the statistical error.  The statistical errors are
estimated by taking the difference between results found with $N$ orbits
(the final number) and $N/2$ orbits, or sometimes --- if that difference
looks suspiciously small --- one half the difference between $N$ orbits and
$N/4$ orbits.  That gives a rough estimate of the error level.  The
statistical errors decrease at large $N$ as $(\ln N)^d/N$ when quasi-random
numbers are used, where $d$ is the number of numbers picked for each
experiment (5 in general, 4 if the binary is circular). That is faster than
the $N^{-1/2}$ decrease that occurs with random numbers (Press et al.\
1992).


The number of orbits needed to reduce the statistical errors to an
acceptable level varies widely with the binary and the measurement.
Measurements are more difficult for $K_1$ than for $H_1$ because of
cancellation.  Cancellation is a problem for $H_1$ too at high $v$ values.
And regardless of what is being measured, binaries with $m_1\gg m_2$ require
many more orbits than equal-mass binaries because of the rare close
encounters with the mass $m_2$.  When the results are presented the number
of orbits used will not be given because it is different for each
measurement; some estimate of the statistical error will be given instead.
The numbers are about $10^4$--$10^5$ orbits per measurement for $H_1$
(sometimes more at high $v$ values) and $10^5$--$10^6$ for $K_1$. The
complete set of experiments took about four months of cpu time on an IBM 580
RISC workstation.

\section{Results from the scattering experiments}

\subsection{Hardening}

\subsubsection{Hardening rate}

\begin{figure*} 
\centerline{\psfig{figure=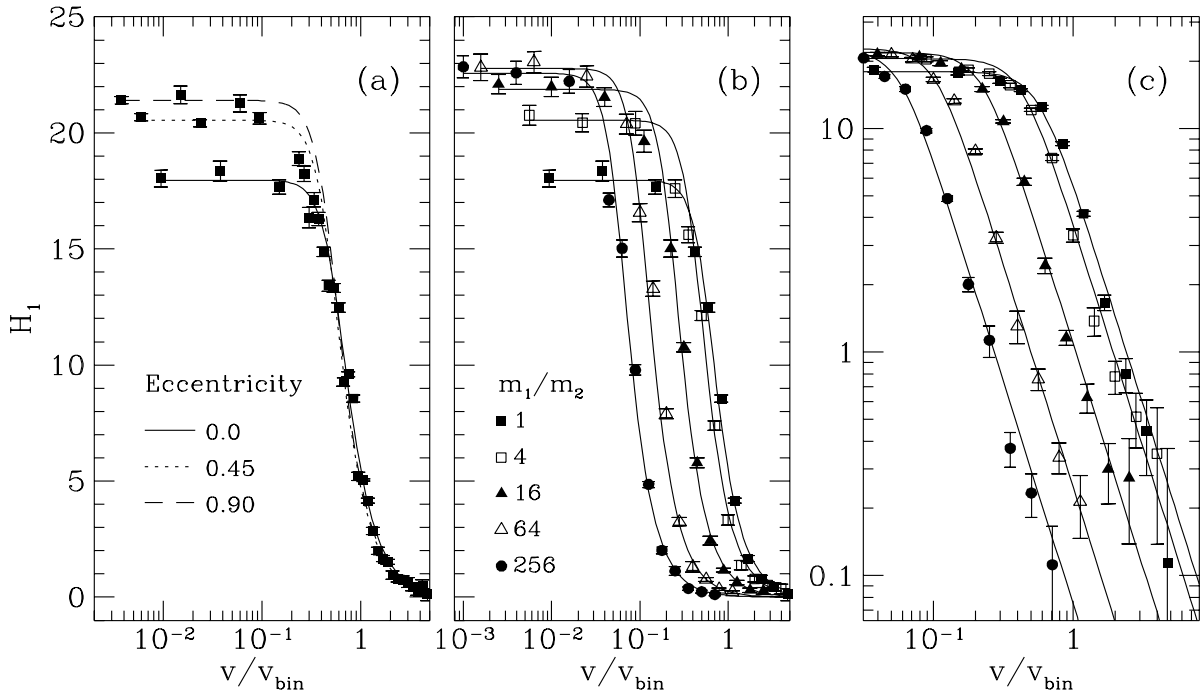,width=\the\hsize}}
\caption[Binary hardening rate.]{Binary hardening rate versus the initial
stellar velocity $v$: (a) shows three eccentricities for $m_1/m_2=1$; (b)
and (c) show five mass ratios for $e=0$, on linear and logarithmic scales.
The lines are the fits to eq.~(\ref{eq-hfit}).}
\label{fig-H1}
\end{figure*}

The hardening rate $H_1$ (\ref{eq-H1def}) has been measured as a function of
the binary's eccentricity and hardness for a wide range of mass ratios.  It
is plotted in Figure~\ref{fig-H1} versus the hardness as given by the ratio
of the initial stellar velocity $v$ and the binary's orbital velocity
$\vbin$ (the relative velocity of the two BHs if the binary is circular):
\begin{equation} \label{eq-vbin}
   \vbin = \sqrt{G\M12/a}.
\end{equation}
The error bars show the statistical-error estimates; if not visible then
they are smaller than the size of the points.

The velocity dependence of $H_1$ is fit by a function with two free
parameters (whose values are given in Table~\ref{tab-Hfit}):
\begin{equation} \label{eq-hfit}
   H_1 = {H_0 \over \left[ 1 + (v/w)^4\right]^{1/2}}.
\end{equation}
The function has a constant value $H_0$ at $v\ll w$, starts to decrease as
$v$ approaches $w$, and decreases as $1/v^2$ at $v\gg w$.  This fits the
data well at high and low velocities. It does not fit so well when $H_1$
first starts to decrease; those data points were given little weight in the
fitting procedure.  A three-parameter function was also tried,
$H_1=H_0/\left[1+(v/w)^\kappa\right]^{2/\kappa}$, but the exponent $\kappa$
never differed by much from 4.

\begin{table}
\begin{center} 
\begin{tabular}{|r|r|r|}
\hline
$m_1/m_2$ & $H_0$ & $w/\vbin$ \\
\hline
  1 & 17.97 &  0.5675 \\
  4 & 20.54 &  0.4263 \\
 16 & 21.87 &  0.2228 \\
 64 & 22.78 &  0.1043 \\
256 & 22.57 &  0.0573 \\
\hline
\end{tabular}
\end{center} 
\caption{Parameters for fits to $H_1$ (eq.~\ref{eq-hfit}) for a circular
binary.}
\label{tab-Hfit}
\end{table}

For an equal-mass binary the constant hardening rate at low velocity,
$H_0\simeq18.0$, agrees well with Mikkola and Valtonen's (1992) $\pi
R_a\simeq18.2$, and also with the results of Hills (1983a, 1992).  At high
velocity $H_1\approx H_0(w/v)^2\simeq 5.8(\vbin/v)^2$, which can be compared
with Gould's (1991) analysis using the impulse approximation, $H_{1,{\rm
Gould}}= (8\pi/3)(\vbin/v)^2$.  The agreement is satisfactory considering
that the error bars are large at high velocity and that the impulse
approximation is justified only if $v/\vbin\gg1$. In fact it is surprising
how well the impulse approximation works when $v/\vbin\simeq 1$.

Panel~(a) shows that the hardening rate for an equal-mass binary does not
vary much with the eccentricity.  Mikkola and Valtonen (1992) reached the
same conclusion, which remains true for all mass ratios.  For later
applications the hardening rate for a circular binary will be used for all
eccentricities because the variation of $H_0$ and $w$ with eccentricity is
too small to matter.

Roos (1981) and Hills (1983a) showed that the low-velocity hardening rate
$H_0$ does not vary much with the mass ratio.  But the velocity $w$ does.
The variation is fit by
\begin{equation} \label{eq-wfit}
   w \simeq 0.85\sqrt{m_2\over \M12}\vbin = 0.85\sqrt{Gm_2\over a} .
\end{equation}
The physical significance of this mass dependence is the following: if $v<w$
the binary can easily capture stars into bound orbits; if $v>w$ it cannot.

The integral~(\ref{eq-Hmb}) for a Maxwellian distribution was evaluated
numerically. The relation between $H$ and $H_1$ is fit closely by the
formula
\begin{equation} \label{eq-Hmbfit}
   {H(\sigma)\over H_1(\sqrt{3}\sigma)} \simeq \sqrt{2\over\pi} + 
   \ln\left[1 + \alpha\left(\sigma\over w\right)^\beta\right],
\end{equation}
with $\alpha=1.16$ and $\beta=2.40$. In the limit of high velocity this
gives
\begin{equation} \label{eq-Hsoft}
   H \approx {\beta H_0\over 6}\left(w^2 \over \sigma^2\right)
       \ln\left(\sigma^2 \over w^2\right) .
\end{equation}
The log term looks like a familiar Coulomb logarithm but comes from an
integral over the velocity distribution, not over a range of impact
parameters. The limit~(\ref{eq-Hsoft}) can be compared with Gould's (1991)
hardening rate,
\begin{equation} \label{eq-Hgould}
   H_{\rm Gould} = {16\sqrt{2\pi}\over 3}\left(G\mu\over a\sigma^2\right)
   \ln\left(\sigma^2\over\vbin^2\right).
\end{equation}
For an equal-mass binary the coefficients of the log terms differ by about
30\%, which is satisfactory considering the uncertainties mentioned above.
For non-equal masses Gould's log term does not have the correct mass
dependence (Gould did not attempt to compute the log term accurately).

The hardening rate for a massive BH binary is sometimes derived from
Chandrasekhar's dynamical-friction formula (e.g.\ Fukushige et al.\ 1992).
The error in that has been known for many years (Chandrasekhar 1944, Hills
1983a): the distant encounters included in the friction formula do not
perturb the binary's semimajor axis --- they only perturb its center of
mass.  It is an accident that the derivation gives a result like Gould's for
a Maxwellian distribution if a suitable choice is made for the log term: if
the same derivation is used for $H_1$ it gives the nonsense result that (for
$m_1=m_2$) $H_1$ is zero at $v=0$, rises as $H_1\sim v$ for $v<\vbin/2$, and
then drops abruptly back to zero at $v=\vbin/2$ (because only stars moving
slower than the BHs contribute in Chandrasekhar's formula). See Gould (1991)
for further discussion.

The velocity dependence of the hardening rate suggests a new convention for
the use of the word hard. A hard binary is usually defined in one of three
ways.  The first says that a binary with binding energy $E_{\rm b}$ is hard
if $E_{\rm b}\gg m_*\sigma^2$ and soft if $E_{\rm b}\ll m_*\sigma^2$ (p.~534
of Binney and Tremaine 1987). The second says a binary is hard if it grows
harder through interactions with stars and soft if it grows softer (Hut
1983). And the third, which is often stated as a corollary of the first or
second rather than as an independent definition, says a hard binary is one
that ``hardens at a constant rate,'' i.e.\ at a rate $H$ that is independent
of the hardness.  The equivalence of the first two definitions is called
Heggie's law. But neither of those definitions is useful for a massive BH
binary because both are satisfied by almost any pair of massive BHs that is
close enough to be called a binary.

The third definition gives almost the same result as the first two when the
masses are equal (see Hut 1983, or Figure~6.3 of Spitzer 1987) and is far
more useful when the binary is massive.  A BH binary will therefore be
called hard if it hardens at a constant rate, i.e.\ if $\sigma\ll w$ or
equivalently if $\vbin/\sigma \gg (1+m_1/m_2)^{1/2}$.  It is tempting to
call a binary soft if it is not hard, but that is confusing for massive
binaries because there is a wide gap for them between a hard binary in the
sense used here and a soft binary in the familiar sense that ``soft binaries
grow softer.''  In later figures the properties of hard binaries are studied
with scattering experiments using the lowest initial velocity $v$ in
Figure~\ref{fig-H1} for each mass ratio; those velocities are
$\log_{10}(v)=-2.025$, $-2.25$, $-2.6$, $-2.8$, and $-3.0$ for $m_1/m_2=1$,
4, 16, 64, and 256.

The scattering results and Gould's (1991) analysis refute Hills's (1990)
statement that a binary grows harder if $\vbin>\sigma$ and softer if
$\vbin<\sigma$ regardless of the values of $m_*$, $m_1$, and $m_2$.
Although the mean energy change $\Cav$ at zero impact parameter does change
from positive to negative when the stellar velocity is raised from $v<\vbin$
to $v>\vbin$, that sign change disappears when $\Cav$ is averaged over
impact parameter. 

The reason for the hard/not-hard transition at $\sigma=w$ is best explained
after we have examined the cross section for stars to be captured by a
binary, to have close encounters with the binary members, and the
distribution of velocities with which stars are expelled from a binary.

\subsubsection{Capture cross section}

We say that a binary captures an incoming star if the star's orbital radius
passes through more than one minimum.  Almost all captured orbits are
eventually expelled in the three-body problem (there might be a set of
measure zero that remain bound forever), but the star can survive for many
revolutions before that happens.

Previous work has used scattering experiments and approximate methods to
derive capture cross sections.  Hills has used scattering experiments to
study the capture of orbits by very hard, massive binaries (Hills 1983a,
1983b, 1992).  He unfortunately defines capture --- or what he calls
long-term capture --- in a way that depends on his program (he says a
long-term capture occurs if the integration takes more than 150,000 steps).
But he gives helpful information on how the capture probability depends on
the impact parameter, eccentricity, and binary mass ratio.  Pineault and
Duquet (1993) have used the impulse approximation to derive approximate
capture cross sections for massive, circular binaries, for arbitrary mass
ratios and degrees of hardness (they give many relevant references to the
comet literature).  They say their cross sections are accurate to within a
factor of 2--3, although that is not clear because they adjust their
formulas in an ad hoc way --- using Hills's (1983a) results to guide them
--- for hard binaries for which the impulse approximation does not work.

\begin{figure*} 
\centerline{\psfig{figure=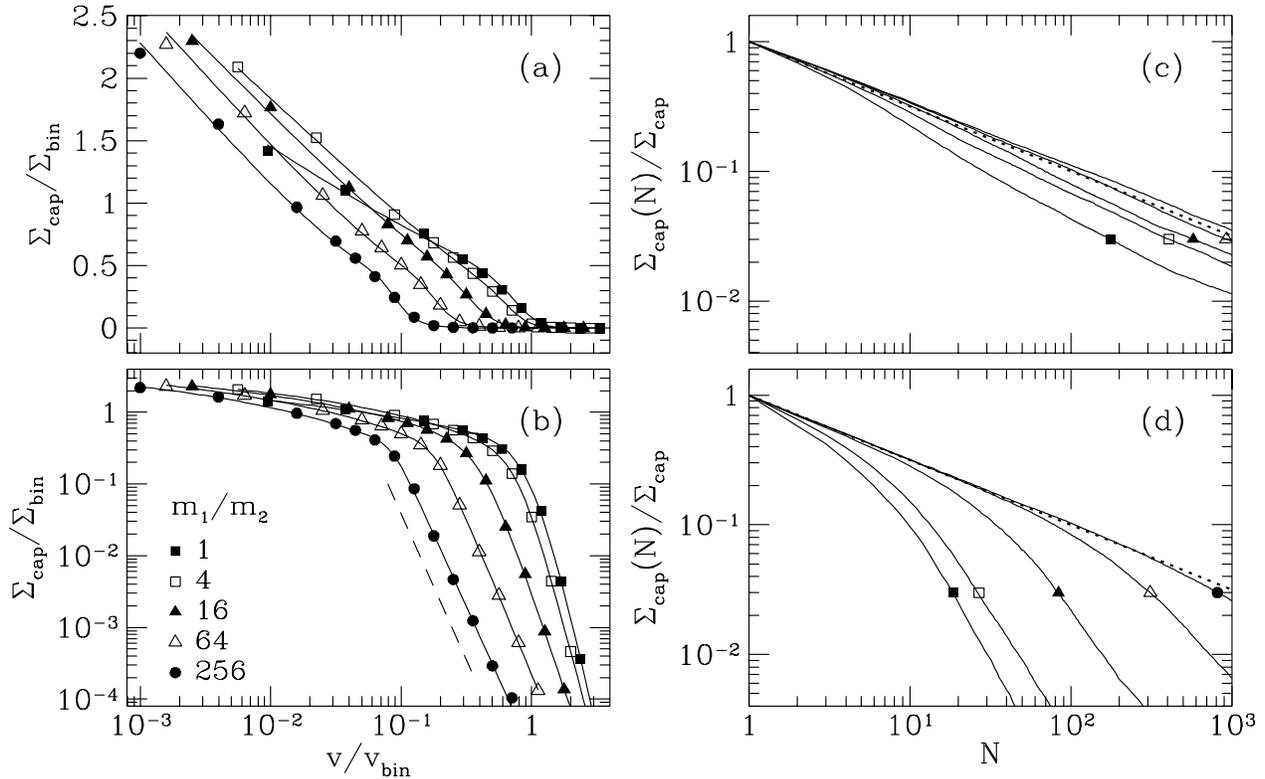,width=\the\hsize}}
\caption[Capture cross section.]{The cross-section $\Sigma_{\rm cap}$ for a 
circular binary to capture an orbit with initial velocity $v$, plotted for
five mass ratios on (a) linear and (b) logarithmic scales.  The solid lines
in (a) and (b) are the fits to eq.~(\ref{eq-Sfit}); the dashed line varies
as $(v/\vbin)^{-4}$. Panels (c) and (d) show the cross section for an orbit
to be captured for at least $N$ revolutions for the same five mass ratios,
plotted in (c) at the lowest $v$ for each mass ratio, and in (d) at $v=w$.
The dotted lines in (c) and (d) vary as $N^{-1/2}$.}
\label{fig-scap}
\end{figure*}

The measurements made here improve upon those of Hills by using a
reproducible capture definition and by exploring the dependence on the
binary's degree of hardness.  Panels (a) and (b) of Figure~\ref{fig-scap}
show the capture cross section for a circular binary in units of the
binary's geometrical cross section $\Sbin$, which includes the correction
for gravitational focusing:
\begin{equation} \label{eq-Sbin}
  \Sbin = \pi a^2\left(1+{2G\M12\over av^2}\right) .
\end{equation}
The velocity dependence is fit by the function
\begin{equation} \label{eq-Sfit}
  {\Scap\over\Sbin} = c_1 
              \left[1+\left(v\over c_2\vbin\right)^{c_3}\right]^{-c_4}
           \ln\left[1+\left(c_2\vbin\over v\right)^{c_5}\right] ;
\end{equation}
the five parameters are listed in Table~\ref{tab-Sfit} (the fits should not
be extrapolated to velocities much higher than shown in the figure).  At low
velocity $\Scap/\Sbin$ rises as $\ln(1/v)$ because the energy change $C$
decreases exponentially with impact parameter. At high velocity
$\Scap/\Sbin$ decreases as a power of $v$, which is clearer for the binaries
with $m_1/m_2\gg1$.  The velocity at the transition between the logarithmic
and power-law behavior, approximately $c_2\vbin$, depends on the mass ratio
in the same way as $w$ (eq.~\ref{eq-wfit}).

For most velocities and mass ratios the cross sections in
Figure~\ref{fig-scap} agree to within a factor of two with the approximate
cross sections of Pineault and Duquet (1993).  There are some larger
differences at high velocity for an equal-mass binary, and at
$v/\vbin\simeq0.01$--$0.1$ when $m_1/m_2\gg1$.  The $(v/\vbin)^{-4}$
behavior seen when $m_1/m_2\gg1$ also agrees with that found by Pineault and
Duquet.

\begin{table}
\begin{center} 
\begin{tabular}{|r|r|r|r|r|r|}
\hline
$m_1/m_2$ & $c_1$ & $c_2$ & $c_3$ & $c_4$ & $c_5$ \\
\hline
  1 & 17.97 & 1.0066 & 3.5745 & 2.0865 & 0.6100 \\
  4 & 20.54 & 0.7929 & 4.5326 & 1.2675 & 1.2377 \\
 16 & 21.87 & 0.4122 & 3.6588 & 1.2324 & 0.9754 \\
 64 & 22.78 & 0.1800 & 6.1855 & 0.5562 & 1.0087 \\
256 & 22.57 & 0.0846 & 8.1992 & 0.3856 & 0.9782 \\
\hline
\end{tabular}
\end{center} 
\caption{Parameters for fits to $\Scap$ (eq.~\ref{eq-Sfit}) for a circular
binary.} 
\label{tab-Sfit}
\end{table}

The capture cross section rises with the binary eccentricity, but the
dependence is weak. The difference in $\Scap$ for circular and
highly-eccentric binaries is only 20--30\%, too small to matter for most
applications.

Panels (c) and (d) of Figure~\ref{fig-scap} show the cross section for a
captured orbit to survive for at least $N$ revolutions, i.e. for the radius
to pass through at least $N+1$ minima before the star is expelled.  The
results for binaries with $m_1/m_2\gg1$ are fit well by $\Scap(N)/\Scap\sim
N^{-1/2}$.  Everhart (1976) noticed this $N^{-1/2}$ scaling in the survival
of comets scattered by the Sun-Jupiter system, and interpreted it as
resulting from a random-walk in the comet's energy, as in the gambler's ruin
problem from probability theory (see Yabushita 1979, Quinn, Tremaine, and
Duncan 1990 for further discussion).  The $N^{-1/2}$ scaling does not work
as well if the binary is not hard or if $m_1\simeq m_2$.

The cross sections in Figure~\ref{fig-scap} place no limit on the apocenter
of the captured orbit. Some of the stars contributing to $\Scap$ are
captured into weakly-bound orbits with apocenters many orders of magnitude
larger than the binary's semimajor axis.  In a real galaxy those orbits will
be perturbed by passing stars and the galactic potential before they return
to the binary. But that should not change the hardening rate much because
the contribution from weakly-bound captures is small, even when $m_1\gg
m_2$.

\subsubsection{Close-encounter cross section}

The cross section for close encounters with the binary members is needed for
applications to real problems where the bodies are not point masses.  For a
massive BH binary we need it to compute the rate at which stars are tidally
disrupted by the BHs, and to estimate how those disruptions might change the
hardening rate.

\begin{figure*} 
\centerline{\psfig{figure=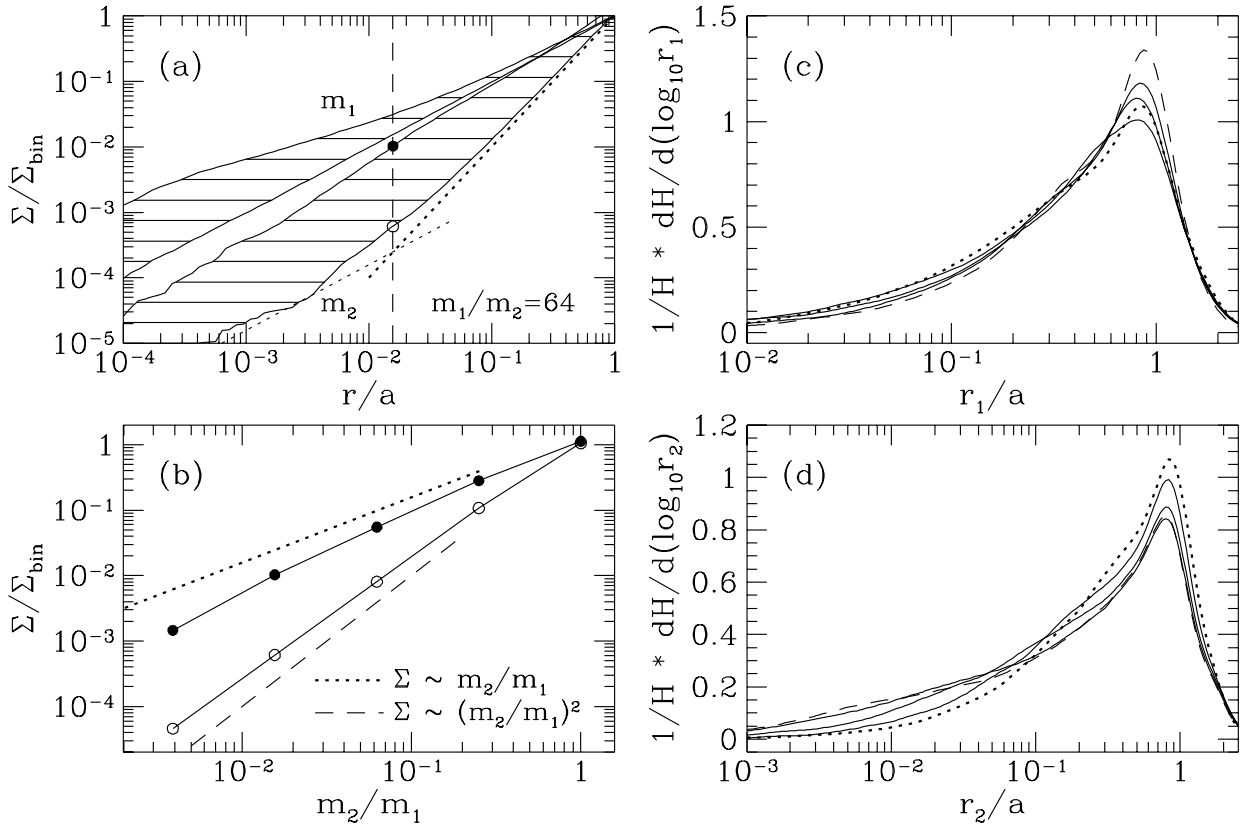,width=\the\hsize}}
\caption[Close-encounter probabilities.]{
Close-encounter cross sections for a hard, circular binary ($v$ equals the
lowest value in Fig.~\ref{fig-H1} for each mass ratio).  Panel (a) shows the
cross section for an orbit to approach within $\leq r$ of $m_1$ or $m_2$.
The lines bounding the upper shaded region are for $m_1$: the lower line
results when the orbits encounter the binary only once; the upper when they
have as many encounters as necessary for them to leave with positive energy.
The lower shaded region has the same meaning, but for $m_2$.  The dotted
lines vary as $r^{-1}$ and $r^{-2}$; the dashed line is at $r/a=m_2/m_1$.
Panel (b) shows the cross section for an orbit to approach within
$r\leq(m_2/m_1)a$ of $m_2$, for both single encounters (open circles) and
multiple encounters (filled circles).  Panels (c) and (d) show the
differential hardening rates with respect to the distances of closest
approach to $m_1$ and $m_2$; the five lines are for $m_1/m_2=1$ (dotted), 4,
16, 64, and 256 (dashed).}
\label{fig-dhdr}
\end{figure*}

Figure~\ref{fig-dhdr}(a) shows the cross section $\Sigma$ for a star to
approach within a distance $\leq r$ of either of the BHs, for a hard,
circular binary with $m_1/m_2=64$.  The cross section is plotted for two
sets of experiments: in the first the stars were allowed to encounter the
binary only once, even if they were captured; the second allowed as many
encounters as necessary for the stars to be expelled.

The cross section for the larger BH scales as $\Sigma/\Sbin\sim r$ for the
single-encounter experiments because of gravitational focusing.  For the
multiple-encounter experiments $\Sigma$ is larger but the increase is
important only for $r/a<m_2/m_1$. The reason the increase is unimportant for
$r/a>m_2/m_1$ is that when a captured star orbits a binary with $m_1\gg m_2$
the star's distance of closest approach to $m_1$ remains nearly constant
while its energy undergoes (approximately) a random walk.  This is well
known in cometary dynamics, where comets diffuse in energy at nearly
constant perihelion (e.g.\ Duncan, Quinn, and Tremaine 1987).

The cross section for a close encounter with the smaller BH is different.
For the single-encounter experiments gravitational focusing is important for
$r/a<m_2/m_1$ but not for $r/a>m_2/m_1$, so $\Sigma/\Sbin$ scales as $r$ or
as $r^2$ depending on whether $r/a$ is smaller or larger than $m_2/m_1$.
For the multiple-encounter experiments the cross section is larger for all
values of $r$, not just for $r/a<m_2/m_1$.

The distance $r/a=m_2/m_1$ has a special importance for $m_2$ if $m_1\gg
m_2$ because the velocity of a star orbiting $m_2$ at that distance equals
$\vbin$.  Figure~\ref{fig-dhdr}(b) shows the cross section $\Sigma$ for such
encounters as a function of the mass ratio. For the single-encounter
experiments $\Sigma/\Sbin\sim (m_2/m_1)^2$; for the multiple-encounter
experiments captures raise that scaling almost to $\Sigma/\Sbin\sim
m_2/m_1$.

Panels (c) and (d) of Figure~\ref{fig-dhdr} show the differential hardening
rates with respect to the distances of closest approach to $m_1$ and $m_2$,
normalized so that the area under the curves is unity. The largest
contribution to the hardening comes from orbits that pass both BHs at a
distance not much smaller than the semimajor axis. When $m_1\gg m_2$ there
is a wide tail in the left of panel (d), but the contribution from close
encounters with $m_2$ is still a small fraction of the total hardening rate.

In a real galaxy there will be two complications that can change these
results. If weakly-bound captured stars are perturbed by nearby stars or the
galactic potential they will not return to the binary in such a way as to
keep their distance of closest approach to $m_1$ nearly constant. That would
increase the difference between the single- and multiple-encounter cross
sections for $m_1$. But if the captured stars are perturbed too much they
might not return at all, which would reduce the cross sections for both
$m_1$ and $m_2$. The two complications tend to cancel for $m_1$.

\subsubsection{Distribution of final velocities}

The final velocity is the velocity of a star at infinity after it has been
expelled by the binary. We need their distribution to compute the mass
ejection rate.

Everhart's (1968, 1969) work on the scattering of comets by planetary
systems is relevant to the distributions to be considered here.  Everhart
used an approximate conic-matching procedure to derive the probability
$h(U)\,dU$ for the energy change $U=\Delta E_*$ to lie in the interval $dU$
after a single encounter between the comet and the planet.  The distribution
has three parts, which Everhart called A, B, and C.  Parts A and B are for
the small and intermediate energy changes and are fit well by (A) a Gaussian
and (B) $h(U)\sim 1/|U|^3$. Part C is for the large energy changes resulting
from rare, close encounters with the planet.  In parts A and B, $h(U)$
depends only on $|U|$, but that symmetry is broken in part C where energy
gains are more frequent than energy losses.

\begin{figure*} 
\centerline{\psfig{figure=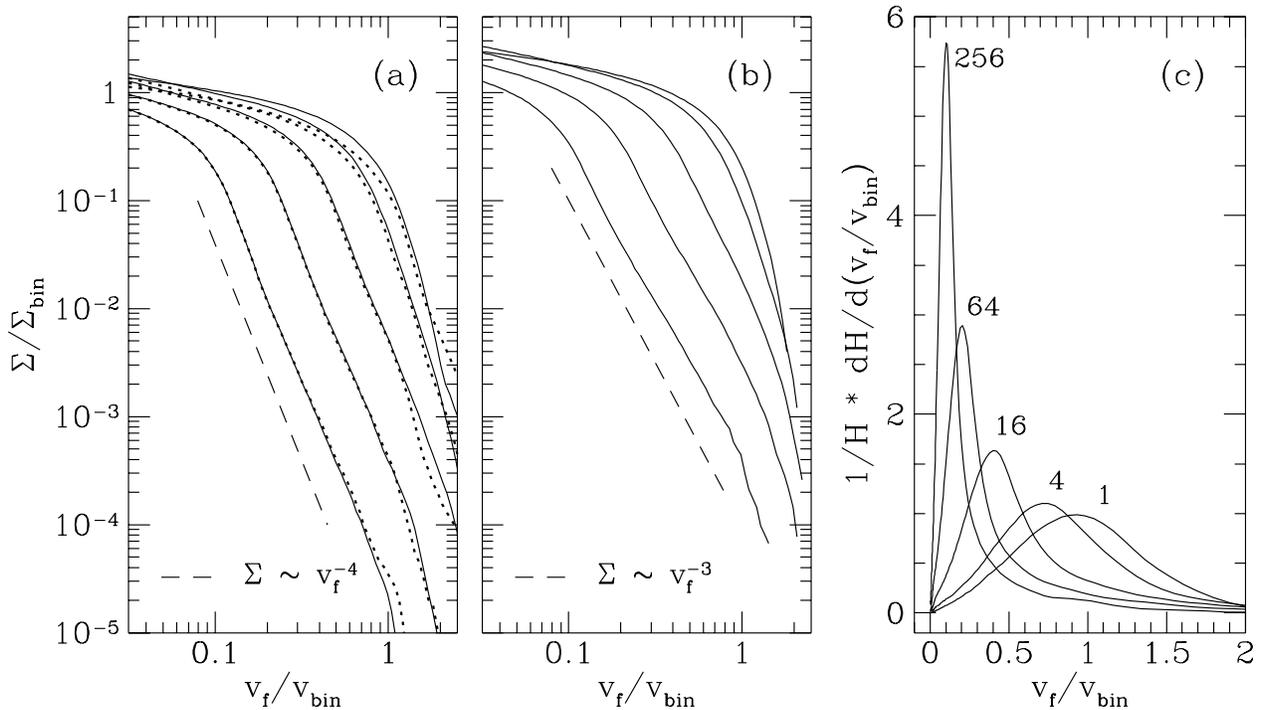,width=\the\hsize}}
\caption[Velocity kicks.]{Distribution of final velocities for a
hard, circular binary ($v$ equals the lowest value in Fig.~\ref{fig-H1} for
each mass ratio).  $\Sigma$ is the cross section for an orbit to leave with
velocity $\geq \vf$ (solid lines), or to remain bound with energy $\leq
-\vf^2/2$ (dotted lines). In (a) the orbits encounter the binary only once;
in (b) as many times as necessary for them to leave with positive energy.
Panel (c) shows the differential hardening rate for the experiments in (b).
The mass ratios $m_1/m_2$ in (a) and (b) are as shown in (c), increasing
from right to left.}
\label{fig-dhdv}
\end{figure*}

Panel (a) of Figure~\ref{fig-dhdv} shows the final-velocity distribution for
a hard binary from scattering experiments done in a manner similar to
Everhart's, so that ``final'' means after a single encounter with the
binary. If a star was captured its final velocity was set to
$-\sqrt{2|E_*|}$, with $E_*$ measured when the star began returning to the
binary for a second encounter; otherwise the final velocity was set to
$\sqrt{2E_*}$ at the end of the integration.  The figure shows the cross
sections $\Sigma$ for the final velocity to be greater than $v_f$ or less
than $-v_f$ for some positive $v_f$.

For the binaries with $m_1/m_2\gg1$ there is a range of velocities for which
$\Sigma$ is symmetric (depends only on $|\vf|$) and varies as $\Sigma\sim
1/v_f^4$. This corresponds to Everhart's part B.  The hardening rate would
be nearly zero for these binaries if multiple encounters were not allowed
because the positive and negative contributions would nearly cancel.  The
symmetry is not as good for binaries with equal masses for which the star is
more likely to gain energy.  That is why the $N^{-1/2}$ scaling in
Figure~\ref{fig-scap} did not work so well for equal-mass binaries.  The
asymmetry that Everhart predicted for part C is not clear in the figure,
perhaps because the statistics are poor when $m_1\gg m_2$ for the rare,
close encounters with the mass $m_2$ (the $m_1/m_2=256$ results come from
$10^6$ orbits, but that is still not enough).

Panel (b) shows the final-velocity distribution when the stars are allowed
to encounter the binary as many times as necessary for them to be expelled.
The $\Sigma\sim 1/v_f^4$ scaling from the single-encounter experiments is
raised to approximately $\Sigma\sim 1/v_f^3$, but the probability of a star
being expelled with $v_f\simeq\vbin$ is still small if $m_1\gg m_2$.

Panel (c) shows the differential hardening rate with respect to the final
velocity, normalized so that the area under the curves is unity. The
velocity at the maximum scales with the mass ratio in the same way as $w$
(eq.~\ref{eq-wfit}) and is approximately $1.75w$.  There is a wide tail to
the right of the maximum if $m_1\gg m_2$ but the high velocities contribute
little to the hardening. In fact the hardening rate for $m_1\gg m_2$ can be
computed quite accurately by considering just the positive velocities from
panel (a) and multiplying the result by two, i.e.\ by assuming that the
captured orbits eventually get expelled with the same distribution of final
velocities as for the orbits that are not captured (this works only for very
hard binaries).

\subsubsection{Discussion}

We can now explain why a hard binary hardens at a constant rate that is
independent of its mass ratio. The explanation given by Roos (1981) is
incorrect. It implies that the dominant contribution to the hardening for a
binary with $m_1\gg m_2$ comes from orbits that have close encounters with
$m_2$ and are expelled with high velocity.  But the scattering experiments
show that the dominant contribution comes from orbits that do not have close
encounters with the BHs and that are expelled with a velocity $v_f\simeq w$.

Consider a typical orbit that starts with a low velocity $v$, passes at a
distance $r\simeq a$ from the two BHs, and leaves with a gain to its kinetic
energy.  The energy gain results mainly from the interaction with the
smaller BH if $m_1\gg m_2$ because the larger BH acts as a fixed potential.
The interaction force of magnitude $F\sim Gm_2/a^2$ acts for a time $\Delta
t\sim (a^3/G\M12)^{1/2}$ to produce a velocity change $\Delta v \sim F\Delta
t \sim (m_2/\M12)\vbin$ and a corresponding energy change $\Delta E_* \sim
\vbin\Delta v \sim (m_2/\M12)\vbin^2$.  That gives $C \sim m_2/\mu \sim 1$,
which is sufficient to give a hardening rate $H_1$ with no dependence on the
hardness and almost no dependence on the mass ratio.

If this same derivation is repeated for a high-velocity star ($v>\vbin$) it
gives a hardening rate that rises as $H_1 \sim v^2$ instead of falling as
$H_1 \sim 1/v^2$ as it should. That is because the derivation ignores the
orbits that lose energy, which tend to cancel the ones that gain energy. The
cancellation removes four powers of $v$ and is the reason the hardening rate
is so difficult to measure at large $v$ by the Monte Carlo method.  For a
hard binary there is no cancellation because the orbits that lose energy in
the first encounter are captured and eventually expelled with an energy
gain. It is not surprising then that the hard/not-hard transition occurs at
the velocity $w$ where the binary begins capturing stars effectively.

\subsection{Mass ejection}

To measure the ejection rate we need an ejection criterion, i.e.\ a velocity
$\vej$ such that a star with initial velocity $v$ is counted as ejected if
it is expelled with final velocity $\vf>\vej$.  The conventional
escape-velocity choice, $\vej = 2\sqrt{3}\sigma$, leads to a problem for a
Maxwellian distribution because 0.7\% of the stars have initial velocities
$v>\vej$ and will be counted as ejected if they receive any energy from the
binary, no matter how little.  We therefore choose $\vej
=\max\{1.5v,2\sqrt{3}\sigma\}$; the results do not depend sensitively on the
numbers 1.5 and $2\sqrt{3}$.  Let $\Fej(x,v,\sigma)$ be the fraction of
stars incident upon the binary with impact parameter $x$ and initial
velocity $v$ that satisfy this criterion.  The ejection rate is then
\begin{equation} \label{eq-Jcalc}
  J = {1\over H}\int_0^\infty\!\! dv\,\,4\pi v^2 f(v,\sigma)\,
      {\sigma\over v} \,4\pi\int_0^\infty\!\! dx\,\,x \Fej(x,v,\sigma) .
\end{equation}
The integral over the velocity distribution is evaluated numerically after
the inner integral is determined from final-velocity distributions like
those shown in Figure~\ref{fig-dhdv}.

\begin{figure} 
\centerline{\psfig{figure=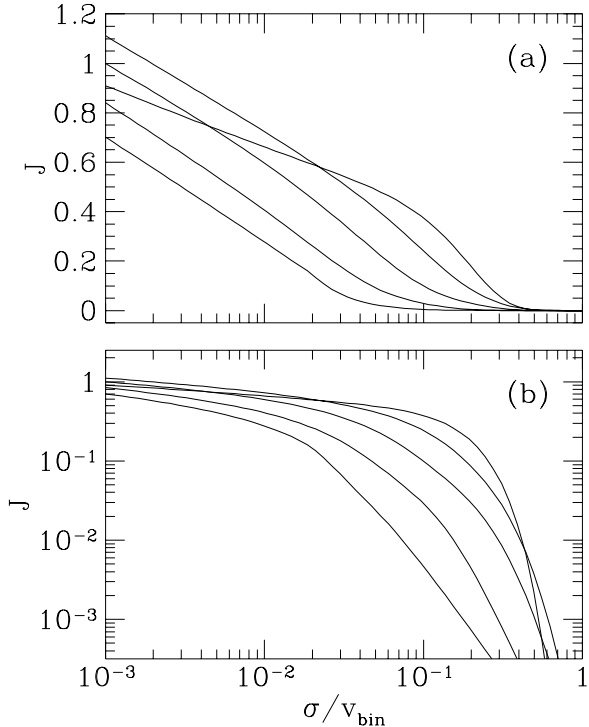,width=\the\hsize}}
\caption[Mass ejection rate $J$.]
{Mass ejection rate $J$ for a circular binary, plotted on (a) linear and (b)
logarithmic scales. The five lines are for mass ratios (increasing from
right to left) $m_1/m_2=1$, 4, 16, 64, and 256.}
\label{fig-J}
\end{figure}

The ejection rate is plotted as a function of $\sigma/\vbin$ for five
mass ratios in Figure~\ref{fig-J}.  At low velocity $J$ rises as
$\ln(1/\sigma)$; at high velocity $J$ falls, first gradually, then
precipitously.  The velocity dependence is fit by the function
\begin{equation} \label{eq-Jfit}
  J = j_1 
              \left[1+\left(v\over j_2\vbin\right)^{j_3}\right]^{-j_4}
           \ln\left[1+\left(j_2\vbin\over v\right)^{j_5}\right] ;
\end{equation}
the five parameters are listed in Table~\ref{tab-Jfit}. The parameters give
a close fit to the data in the figure but are erratic. Note that the
velocity at the bend in the curves in panel~(b) is not fit well by
$j_2\vbin$.

\begin{table}
\begin{center} 
\begin{tabular}{|r|r|r|r|r|r|}
\hline
$m_1/m_2$ & $j_1$ & $j_2$ & $j_3$ & $j_4$ & $j_5$ \\
\hline
  1 & 0.3779 & 0.9200 & 2.2572 & 22.415 & 0.3437 \\
  4 & 0.1148 & 0.8815 & 1.5224 & 10.521 & 1.4162 \\
 16 & 0.0284 & 0.6608 & 0.9404 & 6.7223 & 5.6247 \\
 64 & 0.0665 & 0.4438 & 0.8480 & 8.1901 & 2.1824 \\
256 & 0.2800 & 0.0214 & 3.1294 & 0.5284 & 0.8108 \\
\hline
\end{tabular}
\end{center} 
\caption{Parameters for fits to $J$ (eq.~\ref{eq-Jfit}) for a circular
binary.}
\label{tab-Jfit}
\end{table}

The ejection rate for a hard binary can be estimated by noting that close
encounters with the binary give a mean energy change of $\Cav\simeq 1$.  It
then follows from the definition~(\ref{eq-Cdef}) of $C$ that a binary must
interact with about its own mass in stars to shrink by a factor of $e$.  But
``interact with'' does not mean the same as ``eject.'' A binary that is not
hard interacts with many stars but ejects few of them.  And even a hard
binary need not eject its own mass to shrink by a factor of $e$ if it gives
some stars much more energy than others.

Figure~(\ref{fig-dhdv}) shows that a hard binary with $m_1/m_2\gg 1$ expels
few stars with $\vf>\vbin$. So why does $J$ not decrease rapidly with
$m_1/m_2$ in the left half of Figure~(\ref{fig-J})?  Because although the
fraction of stars expelled with $\vf>\vbin$ does decrease rapidly with
$m_1/m_2$, the fraction expelled with $\vf>w$ does not, and it is that
fraction that determines the ejection rate when the binary first becomes
hard.
 
\subsection{Eccentricity growth}

$K_1$ is more difficult to measure than $H_1$ because of the cancellation
that occurs in the numerator of equation~(\ref{eq-K1}). The $B-C$
distribution is wide and nearly centered on the origin with a mean $\langle
B-C\rangle$ that is 10--100 times smaller than the deviation about the mean.
Consequently we know much less about eccentricity growth than we know about
hardening. Roos (1981) tried to measure $K_1$ with only 500 orbits per
measurement. He found $K_1=0.2\pm0.2$ for a hard binary with $e=0.6$ and
concluded that the eccentricity could increase.  Mikkola and Valtonen (1992)
used $10^4$ orbits per measurement to study the dependence of $K_1$ on the
eccentricity and hardness of an equal-mass binary. Their results are
accurate for hard binaries, for which they found positive growth rates with
a maximum of $K_1 = 0.19 \pm 0.04$, but have large error bars for
$v\simeq\vbin$.

The results derived here use 10 -- 100 times more orbits per measurement
than Mikkola and Valtonen used ($10^5$ per measurement at low stellar
velocity, $10^6$ at high velocity) and use quasi-random numbers rather than
random numbers to further reduce the statistical errors.  The large number
of orbits required has two practical consequences. First, only a small
number of velocities and mass ratios can be examined. Second, the results
should be applied only to problems where $m_*$ is much smaller than $m_1$
and $m_2$, for otherwise the mean behavior can get lost in the dispersion
about the mean (as often happens in N-body experiments).

\begin{figure} 
\centerline{\psfig{figure=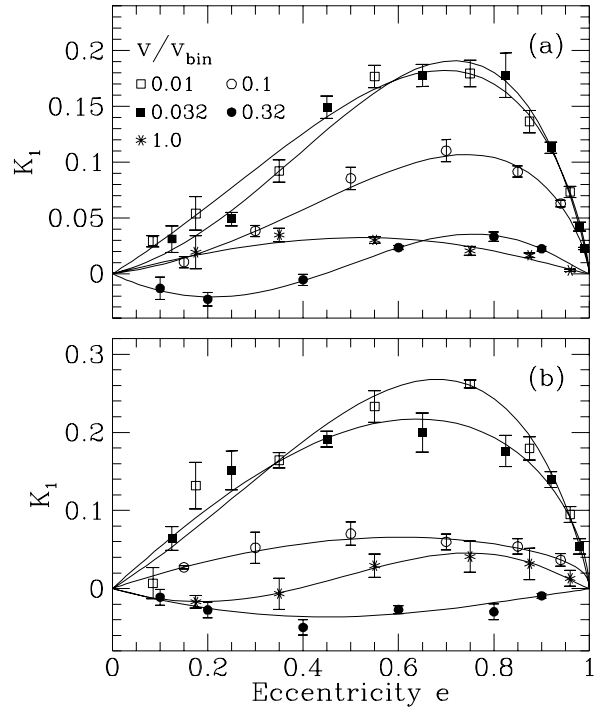,width=\the\hsize}}
\caption[Eccentricity growth rate $K_1$.]
{Eccentricity growth rate for five initial stellar velocities $v$, for
binary mass ratios $m_1/m_2=1$ (a) and 16 (b). The lines are the fits to
eq.~(\ref{eq-k1fit}).}
\label{fig-K1}
\end{figure}

The measurements of $K_1$ are plotted in Figure~\ref{fig-K1} for five
initial velocities and two mass ratios. The eccentricity dependence for each
choice of $v$ and $m_1/m_2$ is fit by the function
\begin{equation} \label{eq-k1fit}
   K_1(e) = e\left(1-e^2\right)^{k_0}\left(k_1+k_2e\right) ;
\end{equation}
the three parameters are listed in Table~\ref{tab-kfit}.  $K_1$ must
approach zero in the limits $e=0$ and $e=1$: the former because of the
conserved Jacobi constant; the latter because of the $(1-e^2)$ in
equation~(\ref{eq-K1}).

\begin{table} 
\begin{center} 
\begin{tabular}{|l||r|r|r||r|r|r||}
\hline
$v/\vbin$ & \multicolumn{3}{c||}{$m_1/m_2=1$}&
\multicolumn{3}{c||}{$m_1/m_2=16$} \\ 
\hline
              & $k_0$ & $k_1$ & $k_2$ & $k_0$ & $k_1$ & $k_2$ \\
\hline
0.01  & 0.731 &   0.265 &   0.230 &  0.822 &   0.383 &  0.402 \\
0.032 & 0.841 &   0.106 &   0.534 &  0.584 &   0.552 & -0.140 \\
0.1   & 0.724 &   0.053 &   0.275 &  0.381 &   0.202 & -0.120 \\
0.32  & 1.271 &  -0.198 &   0.445 &  0.739 &  -0.156 &  0.135 \\
1.00  & 1.169 &   0.102 &  -0.022 &  1.221 &  -0.180 &  0.463 \\
\hline
\end{tabular}
\end{center} 
\caption{Parameters for fits to $K_1$ (eq.~\ref{eq-k1fit}).}
\label{tab-kfit}
\end{table}

Consider first the results for an equal-mass binary.  At large stellar
velocity $K_1$ is small and is negative for some eccentricities.  As the
velocity is lowered $K_1$ rises and becomes positive for all eccentricities.
By $v/\vbin=0.032$, $K_1$ has converged to its limit for a hard binary. That
limit gives a maximum growth rate of $K_1\simeq0.2$ near $e=0.7$, consistent
with the results of Mikkola and Valtonen (1992).  The main difference
between these results and theirs is at high velocity, where they did not
identify negative values for $K_1$. Those values can reduce a binary's
eccentricity when it first starts to harden.

The results for $m_1/m_2=16$ differ from those for an equal-mass binary in
two ways: the maximum $K_1$ is about 40\% larger, and the hard-binary limit
is reached at a lower stellar velocity. In later applications we will need
the growth rate for other mass ratios. We get that by assuming that the
velocity $w$ (eq.~\ref{eq-wfit}) determines the transition to the
hard-binary limit and that this is the only important dependence on the mass
ratio, i.e.\ by interpolating the results for $m_1/m_2=16$ at the
appropriate value of $v/w$.

The derivation of the eccentricity growth rate for a Maxwellian distribution
is more cumbersome than it was for the hardening rate; it is given in an
appendix. The resulting $K$ is the same as $K_1$ at low stellar velocity but
does not fall to zero as fast as $K_1$ at high velocity.  The appendix also
derives $K$ from Chandrasekhar's dynamical-friction formula and shows that
it greatly overestimates the true growth rate.

\section{Application of results to massive black hole binaries}

\subsection{Introduction}

We consider a galaxy core with uniform density $\rho$ and velocity
dispersion $\sigma$.  The core mass and radius, if needed, are computed from
\begin{equation} \label{eq-core}
   \Mc = {4\pi\over 3} \rho \rc^3,\qquad 
   \rc = \left(9\sigma^2\over 4\pi G\rho\right)^{1/2}.
\end{equation}
Model I, for a large galaxy, has $\rho=10^3$, $\sigma=300$,
$\Mc=7.7\times10^9$, and $\rc=120$ (dimensional quantities are given in
units of $M_\odot$, pc, yr, and km/s). Model II, for a small, high-density
galaxy, has $\rho=10^6$, $\sigma=100$, $\Mc=9.0\times10^6$, and $\rc=1.3$.
The BHs are assumed to enter the core and form a binary with initial
eccentricity $e_0$ and semimajor axis $a_0$, the latter chosen so that
$\vbin=\sigma/5$. That is when the integration starts.

The equations for $da/dt$ and $de/dt$ for three-body scattering are combined
with those of Peters (1964) for gravitational radiation:
\begin{eqnarray}                                             \label{eq-adgr}
   \left(da\over dt\right)_{\rm gr} &=&
   -{64\over 5}{ G^3 m_1m_2\M12 \over c^5 a^3 (1-e^2)^{7/2}  }
    \left( 1 + {73\over 24}e^2 + {37\over 96}e^4 \right), \\
                                                             \label{eq-edgr}
   \left(de\over dt\right)_{\rm gr} &=&
   -{304\over 15}{  G^3 m_1m_2\M12 \over c^5 a^4 (1-e^2)^{5/2}  }
    e \left( 1 + {121\over 304}e^2 \right).
\end{eqnarray}
The equation for $d\Mej/dt$ is integrated to give the ejected mass but is
ignored by the other equations, i.e.\ the density and velocity dispersion
are held fixed.

\subsection{Characteristic length and time scales}

It is helpful to first consider how the length and time scales for the
evolution vary with $\rho$, $\sigma$, and the BH masses (see also BBR).  The
binary forms at a separation $\ab = \rc(\M12/\Mc)^{1/3}$ where the enclosed
stellar mass equals $\M12$.  It does not become hard until
$w>\sqrt{3}\sigma$, which happens at about
\begin{equation} \label{eq-ah}
   \ah = {Gm_2\over 4\sigma^2} = 1.2\pc\left(m_2\over 10^8\right)
         \left(300\over \sigma\right)^2.
\end{equation}
{}From then the binary hardens in the time
\begin{eqnarray} \label{eq-th}
   \th &=& \left|a\over \dot a\right| = {\sigma\over G\rho a H} \nonumber \\
       &=& 4.3\times10^7\yr
         \left(\sigma\over 300\right) \left(10^3\over \rho\right)
         \left(0.1\over a\right) \left(16\over H\right), 
\end{eqnarray}
where $H\simeq16$ is a typical hardening rate for a hard binary (recall that
$H$ is $\sqrt{2/\pi}$ times smaller than $H_1$).  This should be compared
with the time for a circular binary to merge through the emission of
gravitational radiation:
\begin{eqnarray} \label{eq-tgr}
   \tgr &=&  {5\over 256}{c^5a^4\over G^3 \mu \M12^2} \nonumber\\
        &=& 2.9\times10^6\yr
         \left(a\over 0.01\right)^4 \left(10^8\over m_1\right)^3
         \left(m_1\over m_2\right) \left(2m_1\over \M12\right) .
\end{eqnarray}
The two are equal at the semimajor axis
\begin{eqnarray} \label{eq-agr}
   \agr = &&\left( {256 \over 5} {G^2 \mu \M12^2 \sigma \over c^5 \rho H}
          \right)^{1/5} = 0.027\pc \left[ \left(m_1\over 10^8\right)^3
         \left(m_2\over m_1\right) \right. \nonumber \\
         &&\left. \left(\M12\over 2m_1\right)
         \left(\sigma\over 300\right) \left(10^3\over \rho\right)
         \left(16\over H\right) \right]^{1/5} .
\end{eqnarray}
The binary orbital velocity at $\agr$ is close to the geometric mean of the
velocity dispersion and the speed of light:
\begin{eqnarray} \label{eq-vgr}
   \vgr &=& \sqrt{G\M12\over \agr} \nonumber\\
        &=& \sqrt{c\sigma}\left[{405\over 16\pi}
         \left(H\over 16\right)    \left(\M12\over \Mc\right)^2
         \left(\M12\over 4\mu\right) \right]^{1/10} ,
\end{eqnarray}
This tells us the hardness at $\agr$, what we previously called
$\sigma/\vbin$:
\begin{equation} \label{eq-grhrd}
   {\sigma\over \vgr} \simeq \sqrt{\sigma\over c} = 
                      0.032 \left(\sigma\over 300\right)^{1/2} .
\end{equation}

Two things should be noted about the length scales. First, $\agr$ is large
enough that tidal disruptions cannot change the hardening rate by much
unless the BHs are very small. The disruption radius about the larger BH for
a star like our Sun is
\begin{equation} \label{eq-rt1}
  r_{\rm t}= R_*\left(m_1\over m_*\right)^{1/3} 
          \!\!\! = 1.0\times 10^{-5}\pc \left(R_*\over R_\odot\right)
               \left(m_1\over 10^8m_*\right)^{1/3} \!\!\! ,
\end{equation}
which gives $r_{\rm t}/\agr\simeq 4.4\times10^{-4}$ for the numbers in
equation~(\ref{eq-agr}).  That is too small to matter because the main
contribution to hardening comes from orbits that do not have close
encounters with the BH.  Accretion of the disrupted stars will not matter
either (although accretion from some larger gas reservoir can make a
difference when the BHs are close).  Tidal disruptions by the smaller BH can
suppress the high-velocity ejections if $m_1/m_2\gg1$, but those ejections
do not contribute much to the hardening.

The second thing to note is the ratio $\ah/\agr$, which determines how many
$e$-foldings the binary has to harden between the time it becomes hard and
the time radiation takes over:
\begin{eqnarray} \label{eq-arat}
  {\ah\over\agr} = 44.0 &&\left[ \left(m_1\over 10^8\right)^2
         \left(m_2\over m_1\right)^4  \left(2m_1\over \M12\right)
         \left(300\over\sigma\right)^{11} \right. \nonumber\\
         &&\left.  \left(\rho\over 10^3\right) 
         \left(H\over 16\right) \right]^{1/5}.
\end{eqnarray}
The larger $\ah/\agr$, the more mass the binary ejects and the more its
eccentricity grows.  For this example $\ah/\agr\simeq\exp(3.8)$ if
$m_1=m_2$, but the ratio is smaller if $m_1$ is small or if $m_2\ll m_1$.

\subsection{Evolution in a fixed galaxy}

\begin{figure*} 
\centerline{\psfig{figure=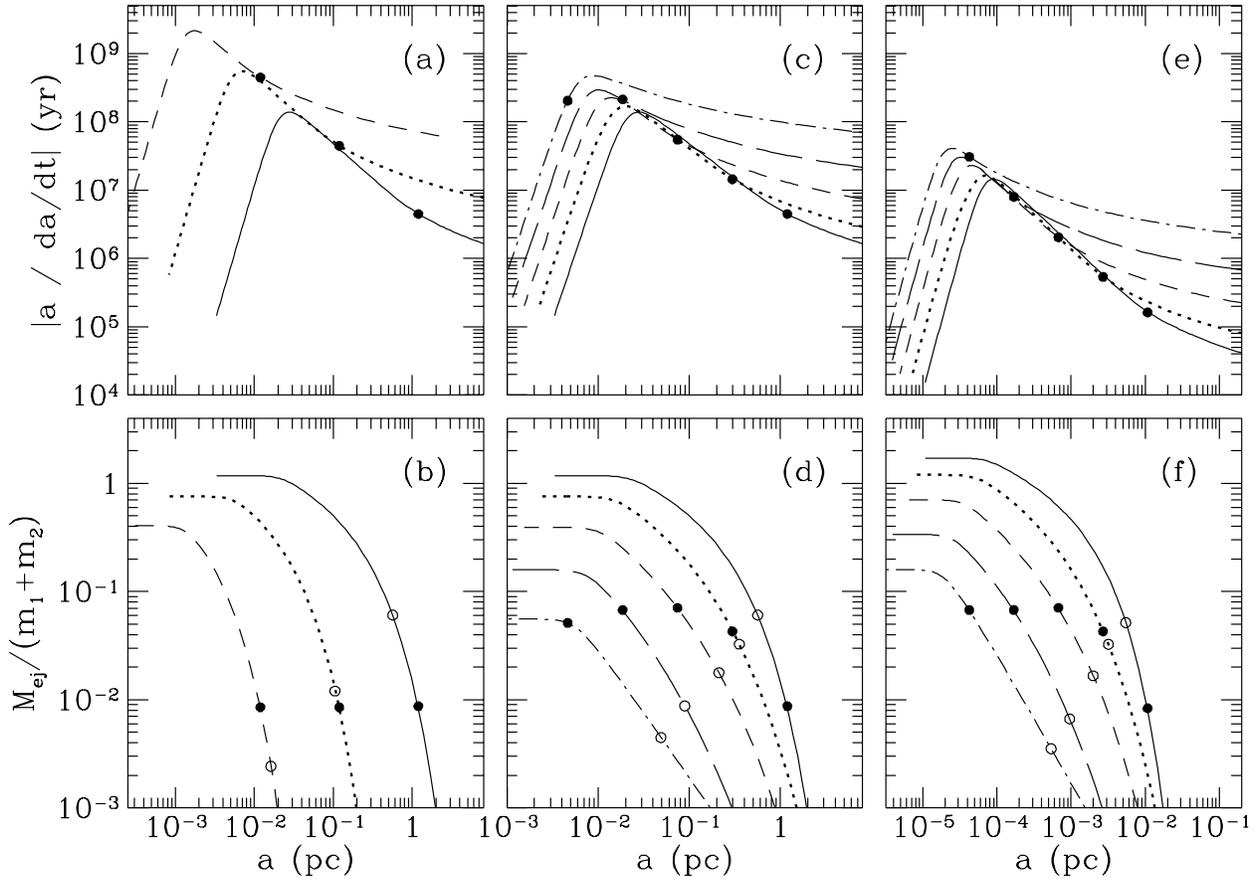,width=\the\hsize}}
\caption[Evolution of (circular) binary BHs.]
{Hardening time $|a/\dot a|$ and ejected mass $\Mej$ versus semimajor axis
for BH binaries in two galaxy models: in model I (panels a--d), $\rho=10^3$
and $\sigma=300$; in model II (panels e and f), $\rho=10^6$ and
$\sigma=100$.  The initial eccentricity is 0.1.  In (a) and (b),
$m_1=m_2=10^8$ (solid line), $10^7$ (dotted), and $10^6$ (dashed).  In (c)
and (d), $m_1=10^8$ and $m_1/m_2=1$ (solid), 4, 16, 64, and 256
(dashed-dotted).  Panels (e) and (f) are similar to (c) and (d) but use
galaxy model II with $m_1=10^5$.  The filled and open circles mark the
points where $a=\ah$ (filled) and where the evolution could stall because of
loss-cone depletion (open).}
\label{fig-bbr6}
\end{figure*}

The equations have been integrated for a number of mass combinations in the
two galaxy models. Figure~\ref{fig-bbr6} shows the hardening time and the
ejected mass for an initial eccentricity of $e_0=0.1$ (the eccentricity does
not grow by much for this choice of $e_0$).  

The hardening proceeds through three stages: the first ends when the binary
becomes hard at $a=\ah$, the second starts there and ends when gravitational
radiation takes over at $a=\agr$, and the third is the final merger stage.
The gradual increase in the hardening time during the first stage is because
of the log term in equation~(\ref{eq-Hsoft}).  All mass combinations harden
along the same diagonal line during the second stage because all hard
binaries harden at the same rate.  The elapsed time equals the hardening
time on that diagonal.  The merger time in the final stage increases as
$m_1$ decreases.  The separation between the stages varies with the masses
as expected: the smaller $m_1$, or the smaller $m_2/m_1$, the less room the
binary has to harden between $\ah$ and $\agr$. Some binaries reach $\agr$
before they become hard.

The hardening curves in panels (a) and (b) are similar to that shown in
Figure~1 of BBR. The main differences are that BBR used a constant hardening
time from Chandrasekhar's dynamical-friction formula during the first stage,
and assumed that all binaries become hard at $\vbin=\sigma$.  The hardening
time used here must match onto Chandrasekhar's time at a large separation,
but not until the log term in equation~(\ref{eq-Hsoft}) matches the usual
Coulomb logarithm.

The ejected mass is never more than a few times $\M12$ and is less if the
BHs are small or if $m_1/m_2\gg 1$. It can be estimated by
$\Jgr\ln(\ah/\agr)/2$ if $\ah/\agr\gg 1$.  The mass ejected by a binary with
$m_1/m_2\gg1$ can be much larger than $m_2$ even though it is not for an
equal-mass binary; so a BH of mass $m$ ejects more mass if it merges with
$N$ BHs of mass $m/N$ than if it merges with another BH of mass $m$.  The
mass displaced from the center of a real galaxy that starts with a density
cusp will be larger than the $\Mej$ computed here, because some is displaced
when the BHs first approach the center and some is nearly but not quite
ejected.

\begin{figure} 
\centerline{\psfig{figure=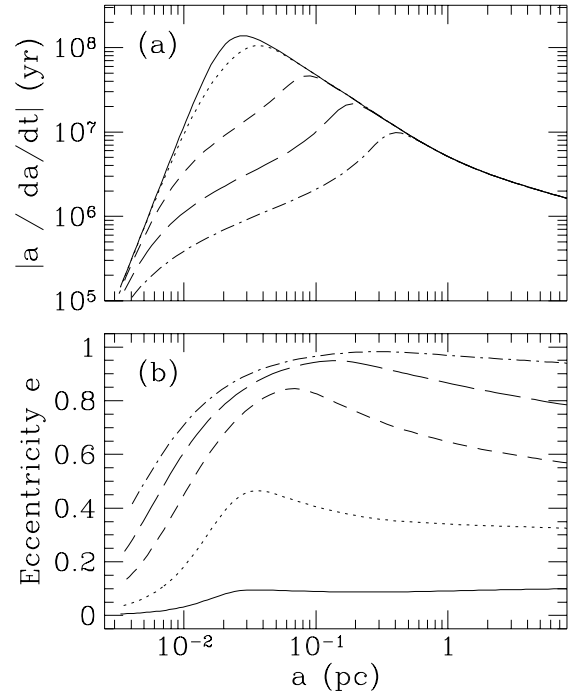,width=\the\hsize}}
\caption[Evolution of eccentric binary BHs.]
{Eccentricity evolution. The solid line is the same as in
Fig.~\ref{fig-bbr6}(a); the other lines assume initial eccentricities of 0.3
(dotted), 0.5, 0.7, and 0.9 (dashed-dotted).}
\label{fig-bbre}
\end{figure}

The dependence on the initial eccentricity is shown in Figure~\ref{fig-bbre}
for one of the binaries from Figure~\ref{fig-bbr6}.  If $e_0\leq0.3$ the
eccentricity hardly grows before gravitational radiation drives it to zero;
if $e_0\simeq 0.1$ the eccentricity goes down from the start.  If $e_0>0.3$
the eccentricity grows, but not by much. This conclusion, which Mikkola and
Valtonen (1992) also reached for an equal-mass binary, is true for all the
binaries in Figure~\ref{fig-bbr6}.  The merger time is reduced if $e_0$ is
large, by about a factor of 20 if $e_0=0.9$ for the binary in
Figure~\ref{fig-bbre}.  The reduction could be larger if the hardening
stalls during the late stages.  Makino et al.\ (1993) say their N-body
experiments show that massive BH binaries form with large eccentricities,
but those experiments use unrealistic galaxy models and atypical initial
conditions.  Polnarev and Rees (1994) give arguments for why the initial
eccentricity should be small.

\subsection{Changes for a galaxy that is not fixed}

We can estimate when mass ejection is likely to affect the evolution by
considering the mass of stars in the unperturbed galaxy that pass within a
distance $r$ of a point mass $\M12$ at the center:
\begin{equation} \label{eq-mrp}
   M(r_p\leq r) \simeq 10\Mc\left(r\over \rc\right)^2
                \left(1+{\M12\over \Mc}{\rc\over r}\right).
\end{equation}
As the binary hardens, $\Mej$ rises and $M(r_p<a)$ falls.  If those two
masses become equal the binary will have ejected nearly all the stars that
can have a close encounter with it; further hardening must then wait for new
stars to diffuse back into the ``loss-cone'' orbits by two-body relaxation.
The point where $\Mej=M(r_p<a)$, marked by the open circles in
Figure~\ref{fig-bbr6}, typically occurs before the binary has ejected even
one tenth of the mass that it has to eject to merge.  The hardening rate
will be reduced by mass ejection before this point is reached.  That will
not change the $\Mej(a)$ and $e(a)$ relations but it will change the time
scale for $a(t)$, perhaps making the merger time longer than the age of the
galaxy for some of the binaries in Figure~\ref{fig-bbr6}.

Mass ejection does not bring the hardening to a complete stop even in the
absence of two-body relaxation because the binary does not remain fixed at
$r=0$. A single particle of mass $\M12$ would wander from the center of the
unperturbed galaxy with an amplitude
\begin{equation} \label{eq-rw}
   \rw \simeq \rc\sqrt{m_*\over\M12} = 
       0.01\pc\left(\rc\over 120\right)
           \left(10^8m_*\over\M12\right)^{1/2}.
\end{equation}
The mass scaling suggests that wandering will be more important for small
BHs, but a binary with large BHs can wander too once it ejects stars from
the center because there is then no restoring force to keep it fixed. The
importance of wandering for the hardening rate is best studied by N-body
experiments; all we can say here is that the merger times in
Figure~\ref{fig-bbr6} are undoubtably too short.

In summary, our study of the restricted three-body problem has answered some
of the questions posed at the start.  It has given a complete description of
how the hardening, mass ejection, and eccentricity growth depend on the
properties of a massive BH binary and a uniform galaxy core in which it is
embedded.  It has cleared up some confusion resulting from mistaken
applications of Chandrasekhar's dynamical-friction formula.  And it has
allowed us to study binary evolution in uniform galaxy cores with some
simplifying assumptions.  But it has left two big questions that affect the
merger time: what the initial eccentricity is and by how much loss-cone
depletion reduces the hardening rate. It has not allowed us to study binary
evolution in realistic galaxy models with density cusps. And it has given
only a crude estimate of the changes induced in such models by the
evolution.  These questions will be the subject of paper II.

\section*{Acknowledgements}

I thank Scott Tremaine for telling me about Everhart's work, and Lars
Hernquist for his encouragement and the use of his workstations.
Financially support was received from NSF grant ASC 93-18185 at UCSC, and
from NSF grant AST 93-18617 and NASA Theory grant NAG 5-2803 at Rutgers.



\appendix
\section{Eccentricity growth rate for a Maxwellian distribution}

The derivation of $K$ from $K_1$ is more cumbersome than the derivation of
$H$ from $H_1$ because $K_1$ is a function of two variables ($e$ and $v$),
because we do not have as much information on its velocity dependence, and
because the energy and angular momentum changes in equation~(\ref{eq-deda})
must be averaged separately. We proceed as follows.  We ignore the
eccentricity dependence of $I_x(C)$, which we know is small, and use
formula~(\ref{eq-hfit}) for the velocity dependence.  The value of $I_x(B)$
is found at any eccentricity and velocity from the assumed $I_x(C)$ and from
linear interpolation of the measured $K_1(e,v)$.  The quantities $I_x(C)/v$
and $I_x(B)/v$ are averaged over the Maxwellian distribution and substituted
into equation~(\ref{eq-K1}) to give the growth rate $K$.  There is some
arbitrariness at high velocity, where we set $K_1(e,v)=0$ if $v>2\vbin$, but
that does not matter for our applications.

\begin{figure*} 
\centerline{\psfig{figure=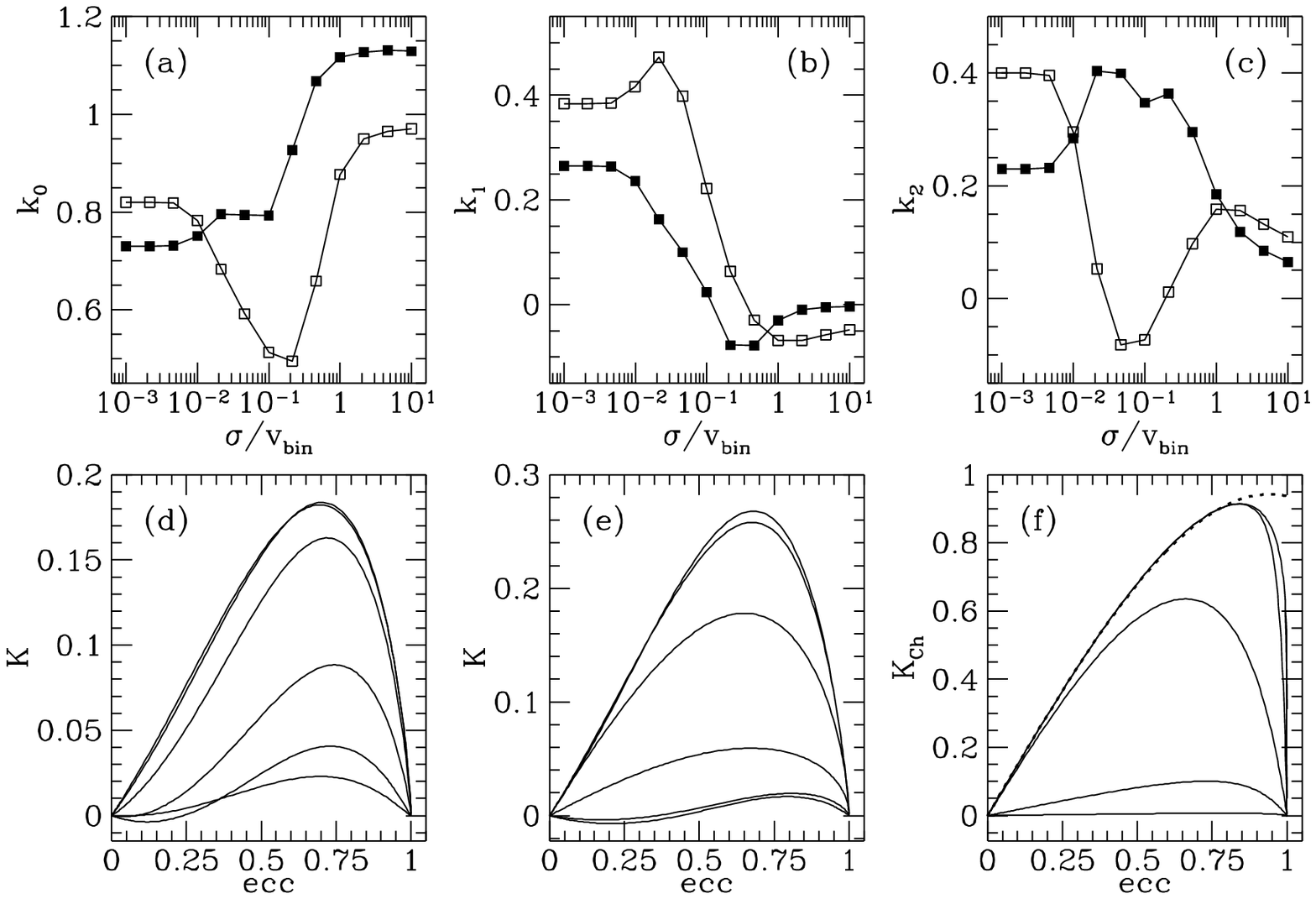,width=\the\hsize}}
\caption[Maxwellian-averaged eccentricity growth rate.]
{Eccentricity growth rate $K$ for a Maxwellian distribution. The top three
panels show the fitting parameters (eq.\ref{eq-k1fit}) for $m_1/m_2=1$
(filled squares) and 16 (open squares).  The bottom three panels show the
fit for $m_1/m_2=1$ (d) and 16 (e), and the prediction (for $m_1/m_2=1$)
from Chandrasekhar's dynamical-friction formula (f); the six lines are for
(from top to bottom) $\sigma/\vbin=0.0025$, 0.01, 0.04, 0.16, 0.64, and 2.56
(the top two lines are not distinguishable in (f)). The dotted line in (f)
is the cubic approximation~(\ref{eq-Kch}).}
\label{fig-K}
\end{figure*}

The growth rate was computed in this way at 13 values of $\sigma/\vbin$ and
fit by the function~(\ref{eq-k1fit}). The parameters for the fits are
plotted in panels (a), (b), and (c) of Figure~\ref{fig-K}; the resulting
fits are plotted in panels (d) and (e) for the two mass ratios.  $K$ and
$K_1$ are about the same for a hard binary; there is no $\sqrt{2/\pi}$
difference as there was for $H$.  $K$ does not fall to zero as fast as $K_1$
at large stellar velocity, and negative values are less prominent for $K$.

These results can be compared with the prediction from Chandrasekhar's
dynamical-friction formula. The frictional force on a massive particle $M$
moving with velocity $V=\sqrt{2}\sigma X$ is (eq.~7-17 of Binney and
Tremaine 1987) 
\begin{equation} \label{eq-dfric}
   {d\vec V\over dt} = -4\pi G^2 M\rho\ln\Lambda{\vec V\over V^3}
   \left[{\rm erf}(X) - {2X\over\sqrt{\pi}}\exp(-X^2)\right] .
\end{equation}
Consider an equal-mass binary in units where $G = \M12 = a = 1$. We ignore
the factor of $4\pi G^2 M\rho\ln\Lambda$ because it cancels in the ratio of
the energy and angular momentum changes.  Those changes are found by
integrating over the unperturbed orbit (we use $V$ here for the relative
velocity of the two BHs):
\begin{equation} \label{eq-del}
   {\Delta E\over E} \propto  \int_0^{2\pi}\!\! dt\,\, 2V^2F(V),\qquad
   {\Delta L\over L} \propto -\int_0^{2\pi}\!\! dt\,\, F(V) ,
\end{equation}
where
\begin{equation} \label{eq-Fv}
   F(V) = {8\over V^3} \left[{\rm erf}\left(X/2\right) - 
          {X\over\sqrt{\pi}}\exp\left(-X^2/4\right)\right] .
\end{equation}
The integrals are evaluated numerically and substituted into
equation~(\ref{eq-deda}) to give the growth rate $\KCh$. Perturbation theory
shows that for a hard binary
\begin{equation} \label{eq-Kch}
   \KCh = {3\over 2}e\left[1 - {3\over 8}e^2 + O(e^4)\right] \qquad
            (X\gg 1),
\end{equation}
which serves as a check on the numerical integration.

The resulting growth rate is plotted in panel~(f). It differs from the
correct growth rate in four ways: it falls to zero too fast as $\sigma$
rises; it rises to the hard-binary limit too fast as $\sigma$ falls; it
approaches zero too slow as $e$ approaches unity; and its maximum value for
a hard binary is about five times too large.  The last three differences
together would greatly reduce the merger time for massive BH binaries if
$\KCh$ were the correct growth rate. But it isn't.

\end{document}